\documentclass[11pt,a4paper,onecolumn]{llncs}
\usepackage{graphicx}
\usepackage{color}
\usepackage{footnote}
\usepackage{epsfig}
\usepackage{amsmath} 

\input{psfig.sty}
\makesavenoteenv{tabular}

\begin{document}

\title{Improving protein threading accuracy via combining local and global potential using TreeCRF model }
\author{Haicang Zhang\inst{1}, Mingfu Shao\inst{1},  Chao Wang\inst{1},  Jianwei Zhu\inst{1},  Wei-Mou Zheng \inst{2}, Dongbo Bu\inst{1,\star}}

\institute{Key lab of intelligent information processing, Institute of Computing Technology, Chinese Academy of Sciences \footnote{ To whom correspondence should be addressed. email: dbu@ict.ac.cn },  \and
 Institute of Theoretical Physics, Chinese Academy of Sciences
}

\maketitle

\begin{abstract}
(Extended Abstract)
\end{abstract}

\section{Motivation}

Protein structure prediction remains to be an open problem in bioinformatics \cite{dill2012protein}. There are two main categories of methods for protein structure prediction: Free Modeling (FM) and Template Based Modeling (TBM). Protein threading, belonging to  the category of template based modeling, identifies the most likely fold with the target by making a sequence-structure alignment between target protein and template protein. Though protein threading has been shown to more be successful for protein structure prediction, it performs poorly for remote homology detection.

Protein residue-residue contacts play critical role in maintaining the proteins\rq{} native structures \cite{gromiha2004inter}. Contacts potential has been used to help improve both  FM and TBM. For FM, the contacts information can help reduce the degrees of freedom in the conformational search space \cite{wu2011improving}\cite{marks2012protein}\cite{michel2014pconsfold}. And for TBM, it can help select the templates sharing similar contact map  with the target protein \cite{ma2015protein}.

Protein threading with contacts potential is NP-hard \cite{lathrop1994protein}. Several approximation algorithms have been proposed to tackle this problem.  PROSPECT proposed divide-and-conquer algorithm to find suboptimal threading alignment \cite{xu2000protein}. RAPTOR formulates the threading problem as an Integer Linear Programming (ILP) and then ILP formulation is  relaxed to a linear programming (LP) problem, which is solved by the canonical branch-and-bound method \cite{xu2003raptor}. MRFalign formulates  the threading problem as a quadratic programming problem and then solve it using Alternate Direction Alternating Direction Method of Multipliers (ADMM) technique \cite{ma2014mrfalign}.

\begin{figure}[!htpb]
	\begin{center}
	\includegraphics[width=1.0 \textwidth]{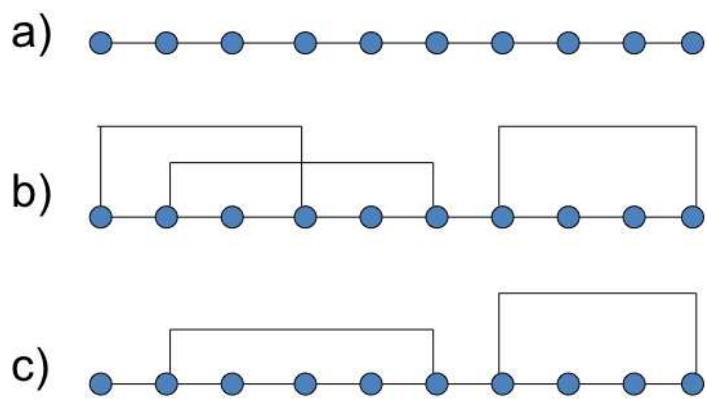}	
	\caption{Graphical models with different structures. a) Chain graph: Inference is easy, but it can\rq{}t  capture global dependence. b) General graph: It can capture global dependence, but inference is hard. c) Nested Graph:  Inference is easy and it can capture global dependence.}
	\label{figure:tree_hardness}
	\end{center}
\end{figure}

  In this paper , we will present our TreeThreader program based on Tree Conditional Random Field (TreeCRF) model. Not only TreeCRF can capture global contact potential, but also the inference in TreeCRF is efficient. In TreeCRF,  the contact pairs of the template are selected to  construct a nested graph.  The special nested structure allows efficient inference  to find the optimum threading alignment. From the view of graphical model, TreeCRF makes  a compromise between model capacity and model complexity. As shown in Figure \ref{figure:tree_hardness},  the inference in ChainCRF is efficient \cite{lafferty2001conditional}, but it can\rq{} capture global dependence. In contrast, CRF with general graph structure can capture global  dependence, but the inference is very hard. The inference in TreeCRF is efficient and it can capture global dependence.


\section{Methods}

Given the template protein and the target protein, the framework of our threading method is as follows.

\begin{enumerate}
	\item Calculate the contact map of the template.
	\item Select the most informative contact pairs of the template using dynamic programming.\\
	\item Prepare the features used in TreeCRF model.\\
	\item Align the target  with the template using TreeCRF model.\\
		
\end{enumerate}

We organize this section as follows. In section \ref{sec:nested}, we will give the dynamic programming algorithm for selecting the most informative contact pairs of the template. Then in section \ref{sec:treecrf}, we will describe our treeCRF model and the details of the inference algorithm. In section \ref{sec:features}, we will describe the alignment  features used in TreeCRF.\\

\subsection{Select the most informative contact pairs} \label{sec:nested}

Given a contact map $G=(V, E)$,  we select the most informative contact pairs by solving the following optimization problem.

\begin{equation}
 \max_{\substack{\quad G'=(V, E')\\ E' \subseteq E \\  G' \text{ is nested}}}  \sum_{(i,j) \in E'} C(i,j) 
 \label{eq:nested}
\end{equation}
 
Here, $C(i,j)$ means the contact potential measuring the importance of the contact pair $(i, j)$. 
Two kinds of contact potential are used in our method: 1) Mutual information (MI) between the sequence profiles of the two residues. 2) Liang-potential \cite{li2003simplicial}.  

We solve the optimization problem \ref{eq:nested} using the following dynamic programming algorithm.

\begin{equation}
M(i,j) = \max \begin{cases}
M(i+1, j-1) + C(i,j)\\
M(i, j-1)\\
M(i-1, j)\\
\max_{i < k < j} M(i,k) + M(k+1,j)
\end{cases}•
\end{equation}•
Here, $M(i,j)$  denotes the optimum from residue the $i$ to the residue $j$. The optimal nested graph can be constructed by the standard traceback procedure of the dynamic programming algorithm.\\

\begin{figure}[!htpb]
	\begin{center}
	\includegraphics[width=1.0 \textwidth]{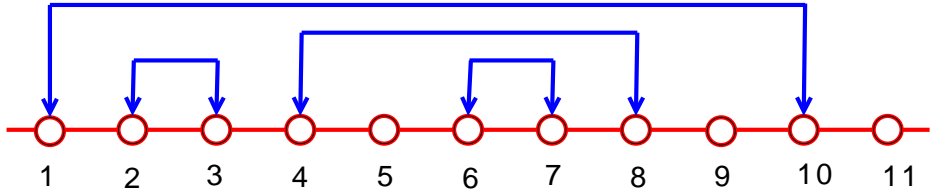}	
	\caption{An example of nested graph}
	\label{figure:nested_example}
	\end{center}
\end{figure}

Each nested graph can be represented by a serial of nodes with different types ($L$, $R$, $P$ and $B$).
Type of the node indicates the direction of the subgraph (left, right, pair and bifurcation).
For example,  the nested graph in figure \ref{figure:nested_example} can be represented as
\{$R(1,11)$, $P(1,10)$, $R(2,9)$, $B(2,8)$, $P(2,3)$, $P(4,8)$, $L(5,7)$, $P(6,7)$\}.

\subsection{TreeCRF model}\label{sec:treecrf}

Let $T$ denote a template protein and $S$ a target protein.
 Each protein is associated with some protein features, such as sequence profile and secondary structure. Let $A=\{a_1,a_2,\cdots,a_L\}$ denote an alignment between $T$ and $S$ where $L$ is the alignment length and $a_i$ is one of the three possible states $M$ (Match), $I_s$ (Insertion), $I_t$ (Delete). In TreeCRF, the probability of an alignment $A$ is calculated as follows.

\begin{equation}
P(A |T,S,\theta)  = \frac{1}{Z(T,S)} \exp\{ \sum_{i=1}^L f(a_{i-1},  a_i, T, S) + \sum_{(i,j) \in E'}  g(a_i, a_j, T, S) \}
\label{eq:alignment_probability}
\end{equation}•
, where  $f$ and $g$ denote local alignment potentials and global alignment potential respectively. We will give the details of these alignment potential in Section \ref{sec:features}. In Eq. \ref{eq:alignment_probability}, $Z_(S,T)$ denotes the partition function calculated as 
\begin{equation}
Z(S,T) = \sum_{\{a_1, \cdots, a_L\}}  \frac{1}{Z(T,S)} \exp\{ \sum_{i=1}^L f(a_{i-1},  a_i, T, S) + \sum_{(i,j) \in E'}  g(a_i, a_j, T, S) \}
\end{equation}•

In ChainCRF or Hidden Markov Model (HMM) \cite{eddy1996hidden}, Forward algorithm and Backward algorithm are used to calculate the partial alignment probability $P(a_1, a_2, \cdots, a_k | T, S, \theta)$ and $P(a_k, a_{k+1}, \cdots, a_L | T, S, \theta)$ respectively. Viterbi algorithm is used to calculated the optimal alignment by maximizing the alignment probability.

\begin{equation}
\max_{a_1, a_2, \cdots, a_L} P(a_1, a_2, \cdots, a_L | S, T, \theta )
\label{eq:max_alignment}
\end{equation}•
All the above three algorithm are standard dynamic programming algorithms with time complexity $O(m^2n^2)$, where $m$ and $n$ are the length of the template protein and the target protein respectively. 

In contrast,  we developed  Outside algorithm and Inside algorithm to calculate the partial alignment probability and  Tree-Viterbi algorithm  to calculate the optimal alignment.

 Let $O(i,j)$ and $I(i,j)$  denote the partial alignment probability $P(a_1, a_2, \dots, a_{i-1}, a_i, a_j, a_{j+1}, \cdots, a_L | T, S, \theta)$ and $P(a_i, a_{i+1}, a_{i+2}, \cdots, a_{j-2}, a_{j-1}, a_j | T, S, \theta)$ 
respectively.  $O(i,j)$ and $I(i,j)$ are calculated recursively as follows.

\begin{equation}
O(i,j) = \sum _{a_{i-1}, a_{j+1}}[ \exp(f(a_{i-1}, a_{i}, T, S) + f(a_{j}, a_{j+1}, T, S) + g(a_i, a_j, T, S) )O(i-1, j+1)]
\label{eq:outside}
\end{equation}•

\begin{equation}
I(i,j) = \sum _{a_{i+1}, a_{j-1}}[ \exp(f(a_i, a_{i+1}, T, S) + f(a_{j-1}, a_j, T, S) + g(a_i, a_j, T, S) )I(i+1, j-1)]
\label{eq:inside}
\end{equation}•

\begin{figure}[!htpb]
	\begin{center}
	\includegraphics[width=1.0 \textwidth]{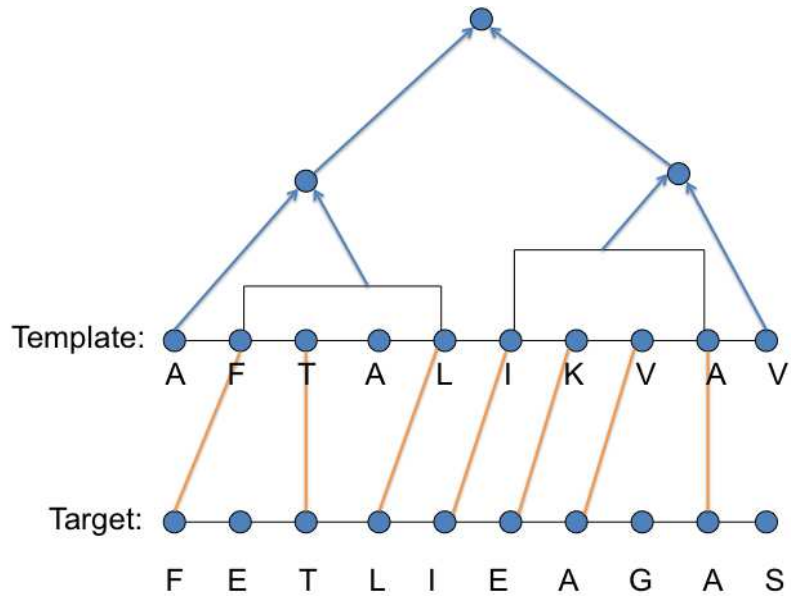}	
	\caption{The process of calculating partial alignment probability using Inside algorithm.}
	\label{figure:tree_alignment}
	\end{center}
\end{figure}

Figure \ref{figure:tree_alignment} shows the process of the Inside algorithm.  The Inside algorithm calculates the partial  alignment from the inside to the outside following the tree structure.

\begin{table}
\centering
\begin{tabular}{ c| c| c }
\hline
   ChainCRF & TreeCRF & CRF with general structure \\
\hline
 Forward algorithm ($O(mn)$)   & Inside algorithm ($O(Kmn)$)  & NP-hard\\
 \hline
 Backward algorithm ($O(mn)$)   & Outside algorithm ($O(Kmn)$)  & NP-hard\\
\hline
 Viterbi algorithm ($O(mn)$)   & Tree-Viterbi algorithm ($O(Kmn)$)  & NP-hard\\
\hline
\end{tabular}
\label{tab:time_complexity}
\caption{The comparison between the complexity of ChainCRF, TreeCRF and CRF with general structure}
\end{table}

The time complexity of Outside algorithm, Inside algorithm and Tree-Viterbi algorithm is $O(Kmn)$, where $K$ is the number of the selected contact pairs of the template. As shown  in Table \ref{tab:time_complexity}, TreeCRF makes a compromise between model capacity and model complexity.

\subsection{Alignment features} \label{sec:features}

The features used to estimate the alignment probability of two residues is as follows.
\begin{enumerate}
\item Sequence profile similarity: the profile similarity between two positions is calculated as \cite{soding2005protein}
\begin{equation}
S_{aa}(q_i, p_j) = \log(\sum_{a=1}^{20}\frac{q_i(a)p_j(a)}{f(a)})
\end{equation}•
Here, $q_i(a)$ and $p_j(a)$ denote the  frequency of amino acid $a$ at the $i$th position of the template and the $i$th position of the target. And $f(a)$ means the background frequency of amino acid $a$.
\item Secondary structure score: we generate 8-class secondary structure types for the template using DSSP \cite{joosten2011series} and predict the 3-class secondary structure types for the target using PSIPRED \cite{mcguffin2000psipred}. The secondary structure  score is calculated as
\begin{equation}
SS(\delta; \rho, c) = \log \frac{P(\delta; \rho, c)}{P(\delta)P(\rho,c)}
\end{equation}
Here, $\delta$ the secondary structure type of the template and $(\rho, c)$ means the secondary structure of the target predicted  as $\rho$ with confidence $c$.

\item Solvent accessibility (SA) score: Real value SA of the query is predicted by Real-SPINE \cite{dor2007real} and  SA of template are calculated by DSSP. The SA score is calculated as: $1-2|sa(i)-sa(j)|$ where $sa(i)$ is the residue solvent accessibility of target sequence predicted by Real-Spine and  $sa(j)$ is  the residue solvent accessibility of the template calculated by DSSP.
	
\item Dihedral torsion angles: The real value torsion angle of the query is predicted by Real-SPINE and that of template is calculated by DSSP. The difference between predicted angles $(\psi(i)$ and $\phi(i))$ of the query  and actual angles $(\psi(j)$ and $\phi(j) )$  of the template is characterized  
\begin{equation}
\bigtriangleup=
	\sqrt{\frac{1}{2}[(\psi(i)-\psi(j))^2 + (\phi(i)-\phi(j))^2]}
\end{equation}
\item Environment fitness score:  This score measures how well one sequence residue aligns to a specific template environment.
\end{enumerate}

\subsection{Results}

We constructed PDB25 dataset using PDB-SELECT \cite{griep2009pdbselect}.  Any two proteins in PDB25 share $< 25\%$ sequence identity. Then we randomly select 300 protein pairs as  training data and another 300 pairs as testing data.  There is no redundancy between the training and testing data . The reference structure alignments for the training and testing data are built using TMalign \cite{zhang2005tm}. 

We compare our TreeCRF threading method, named TreeThreader with the widely used software HHpred \cite{soding2005protein}. As shown in \ref{result:treecrf},  TreeThreader achieves better performance than HHpred.

\begin{table}
\centering
\begin{tabular}{ c|c |c | c | c  }
\hline
  & TM-align & HHpred-mac & Tree-Viterbi & Tree-mac  \\
\hline
GDT& 51.1   & 33.1  & 33.9    & 35.8 \\
\hline
\end{tabular}\\

\caption{Reference-dependent alignment accuracy of TreeCRF and HHpred on test dataset of 300 pairs}
\label{result:treecrf}

\end{table}

\section{Conclusion}

We developed a novel protein threading tool named TreeThreader. Firstly, both local potential and global potential are used in TreeThreader. Secondly,  the TheeThreader is very efficient and practical. Results show that TreeThreader achieves better performance than the widely used protein alignment tool HHpred.

\section*{Acknowledgment}

The study was funded by the National Basic Research Program of China (973 Program) under Grant 2012CB316502, the National Nature Science Foundation of China under Grants 11175224 and 11121403, 31270834, 61272318, 30870572, and 61303161 and the Open Project Program of State Key Laboratory of The- oretical Physics (No.Y4KF171CJ1). This work made use of the eInfrastructure provided by the European Commission co-funded project CHAIN-REDS (GA no 306819).

\bibliographystyle{plain}

\end{document}